\documentclass[a4paper,fleqn,usenatbib]{mnras}

\usepackage{graphicx}
\usepackage{color}
\usepackage{amssymb}
\usepackage{txfonts}

\def\beq{\begin{equation}}
\def\eeq{\end{equation}}
\def\bea{\begin{eqnarray}}
\def\eea{\end{eqnarray}}

\title[The distinguishing signature of Magnetic Penrose Process]{The distinguishing signature of Magnetic Penrose Process}
\author[N. Dadhich et al.]{Naresh Dadhich$^{1}$\thanks{e-mail: \href{mailto:nkd@iucaa.in}{nkd@iucaa.in}},
Arman Tursunov$^{2}$\thanks{e-mail: \href{arman.tursunov@fpf.slu.cz}{arman.tursunov@fpf.slu.cz}}, %\newauthor
Bobomurat Ahmedov$^{3,\;4}$\thanks{e-mail: \href{ahmedov@astrin.uz}{ahmedov@astrin.uz}},
and
Zden\v{e}k Stuchl{\'i}k$^{(2)}$\thanks{e-mail: \href{mailto:zdenek.stuchlik@fpf.slu.cz}{zdenek.stuchlik@fpf.slu.cz}}
\\
$^{1}$IUCAA, Post Bag 4, Ganeshkhind, Pune 411 007, India\\
$^{2}$Institute of Physics and Research Centre of Theoretical Physics and Astrophysics,\\
 Silesian University in Opava, Bezru{\v c}ovo n{\'a}m.13, CZ-74601 Opava, Czech Republic \\
$^3$National University of Uzbekistan, Tashkent 100174, Uzbekistan\\
$^4$Ulugh Begh
Astronomical Institute, Astronomicheskaya 33, Tashkent 100052,
Uzbekistan
}

\date{Accepted XXX. Received YYY; in original form ZZZ}

\pubyear{2018}

\begin{document}

\date{}

\pagerange{\pageref{firstpage}--\pageref{lastpage}} \pubyear{2018}

\maketitle

\label{firstpage}

\begin{abstract}
{In this Letter, we wish to point out that the distinguishing feature of Magnetic Penrose process (MPP) is its super high efficiency exceeding $100\%$ (which was established in mid 1980s for discrete particle accretion) of extraction of rotational energy of a rotating black hole electromagnetically for a magnetic field of milli Gauss order. Another similar process, which is also driven by electromagnetic field, is Blandford-Znajek mechanism (BZ), which could be envisaged as high magnetic field limit MPP as it requires threshold magnetic field of order $10^4$G. Recent simulation studies of fully relativistic magnetohydrodynamic flows have borne out super high efficiency signature of the process for high magnetic field regime; viz BZ. We would like to make a clear prediction that similar simulation studies of MHD flows for low magnetic field regime, where BZ would be inoperative, would also have super efficiency. }
\end{abstract}


%\maketitle

%%%%%%%%%%%%%%%%%%%%%%%%%%%%%%%%%%%%%%%%%%%%%%%%%%%%%%%%%%%%%%%%%%%%%%%%%%%
\section{Introduction}\label{intro}%%%%%%%%%%%%%%%%%%%%%%%%%%%%%%%%%%%%%%%%
%%%%%%%%%%%%%%%%%%%%%%%%%%%%%%%%%%%%%%%%%%%%%%%%%%%%%%%%%%%%%%%%%%%%%%%%%%%

One of the most fascinating aspects of Einstein's theory of gravity (GR) is that black hole mechanics and thermodynamics have uncanny correspondence.
In particular, the second law of thermodynamics of non-decreasing entropy of an isolated system translates to black hole area non-decrease in its interaction with matter fields.
The association of gravity with thermodynamics has come a long way, beginning with formulation of laws black hole mechanics \citep{Bar-Car-Haw:1973:CMP:} to gravity as thermodynamics \citep{Jacobson:1995:PRL:,Padmanabhan:2010:RRP:}.  A black hole is indeed endowed with the thermodynamical properties, entropy and temperature which are respectively proportional to horizon area and surface gravity. Entropy of black hole is given by $S_{\rm BH} = (2\pi c/\hbar) M r_H$ which can only increase with time for an isolated system and remains constant in the limiting case of reversible processes \citep[see, e.g.][]{Mis-Tho-Whe:1973:Gra:}. From this it follows in a straightforward manner that black hole must have an irreducible energy which cannot be transformed into work \citep{Bekenstein:1973:PHYSR4:} Thus the maximum extractable energy from a maximally rotating ($a=M$) black hole, $E_{\rm rot}$, and corresponding irreducible energy, $E_{\rm irr}$, are given by
\beq
E_{\rm rot} = M c^2 - E_{\rm irr}, \,\,\, E_{\rm irr} = \frac{M c^2}{\sqrt{2}}\left[1+\sqrt{1-\left(\frac{a}{M}\right)^2}\right]^{\frac{1}{2}},
\label{ExtRotEn}
\eeq
where $M$ and $a$ are respectively black hole mass and specific angular momentum.
Dividing Eq. (\ref{ExtRotEn}) by total energy $M c^2$, we get the fraction of total energy of a rotating black hole available for extraction.
The energy being extracted is rotational energy of the hole which increases with increasing rotation parameter. One can see from (\ref{ExtRotEn}) that the maximum  extractable energy, which is the measure of rotational energy, is equal to $1-2^{-1/2} \approx 0.29$ or $29\%$ of the black hole energy.

There are two main mechanisms by which rotational energy of a black hole could be extracted out, Penrose process (PP) \citep{Penrose:1969:NCRS:} and its magnetic version (MPP) \citep{Wag-Dhu-Dad:1985:APJ:}, and Blandford-Znajek mechanism (BZ) \citep{Bla-Zna:1977:MNRAS:}. PP exploits purely geometric property that allows existence of negative energy orbits in the vicinity of a rotating black hole while its electromagnetic version (MPP) and BZ in addition to negative energy orbits are also fuelled by quadrupole electric field produced by twisting of magnetic field lines. Both these processes are mediated by magnetic field, and it is this mediation which is responsible for enormous increase in efficiency of energy extraction exceeding $100\%$.

The original mechanical PP was not efficient enough for its astrophysical viability. This was because energy required for a particle to get onto negative energy orbit was quite significant which translated into relative velocity between the fragments to be greater than $1/2 c$ \citep{Bar-Pre-Teu:1972:APJ:,Wald:1974:PHYSR4:,Wald:1974:APJ:}. There is no conceivable astrophysical mechanism that can instantaneously accelerate particles to such a high velocity. This was however beautifully circumvented in MPP \citep{Wag-Dhu-Dad:1985:APJ:} where required energy could now come from electromagnetic field leaving relative velocity to be free. This was how PP was revived in mid 1980s for astrophysical applications as powering engine for high energy sources like quasars and AGNs.  It was shown that MPP was enormously efficient with efficiency even exceeding $100\%$ \citep{Par-etal:1986:APJ:,Bha-Dhu-Dad:1985:JAPA:}. This was however established for discrete particle accretion, and it is gratifying that this prediction of efficiency exceeding $100\%$ has been borne out by recent numerical studies of fully relativistic magnetohydrodynamical fluid flow simulations \citep{Nar-McC-Tch:book:2014:,Tch-Nar-McK:2011a:MNRAS:}.

On the other hand BZ extracts energy from a rotating black hole analogously to the Faraday unipolar generator with the disc (substituted by a black hole) rotating in external magnetic field. The rotation of a black hole generates electric current along its surface which allows the rotational energy of the black hole to be converted into extractable electromagnetic energy. %As in the classical Faraday disc, BZ is quite inefficient as we demonstrate below.
It however requires for a stellar mass black hole threshold magnetic field of order $10^4G$ for creation of force free magnetosphere for it to be operative. In contrast MPP attains super efficient state (efficiency $\gtrsim 100\%$) for a few milli Gauss (mG) field.

 The driving force for both the processes is quadrupole electric field produced by twisting of magnetic field lines. Discharge of which by infalling oppositely charged negative energy flux (for BZ this is how current circuit is completed ) resulting in extracting rotational energy of black hole. Thus the two processes are essentially the same as recent studies have shown \citep{Nar-McC-Tch:book:2014:,Tch-Nar-McK:2011a:MNRAS:} that BZ is high magnetic field limit of MPP. The former requires threshold magnetic field of $10^4G$ for it to be operative while the latter always works, even in its absence when it goes over to original mechanical PP. It is rotational energy of black hole which is being extracted electromagnetically in both the cases. The main aim of this letter is to  clarify the fact that electromagnetic field only serves as catalytic agent while extracted energy is always rotational, and also to emphasize the feature that super high efficiency $>100\%$ is the distinguishing signature of electromagnetic extraction of rotational energy.

Throughout the paper we employ the Kerr geometry in the Boyer-Lindquist coordinates, which is
characterized by two parameters, specific angular momentum $a$ and mass $M$ of the black hole. The physical singularity occurs at the ring $r=0, \theta = \pi/2$. The horizon is given by $r_{H} = M + (M^2-a^2)^{1/2}$, and the static surface by $r_{\rm stat}(\theta) = M + (M^2-a^2 \cos^2\theta)^{1/2}$, defines the outer boundary of ergosphere.
We assume that magnetic field surrounding black hole does not affect the background Kerr metric. This assumption holds when magnetic field strength satisfies the condition $B << c^4 / (G^{3/2} M) \sim 10^{19}{(M_{\odot}/ M)}$G, which is constrained by the equipartition between black hole mass and magnetic field energy \citep{Tur-Stu-Kol:2016:PHYSR4:}. In all known astrophysical phenomena occurring in black hole vicinity, this condition is perfectly satisfied. For example, there are several theoretical and observational results giving the typical magnetic field estimates of order $B \approx 10^8$G for $M\approx 10 M_{\odot}$, and $B \approx 10^4$G for $M\approx 10^9 M_{\odot}$ \citep[see, e.g.][]{Pio-etal:2011:ASBULL:,Baczko-etal:2016:AAP:}.

\section{Magnetic Penrose Process} \label{sectionMPP}

In 1969 Penrose came up with an ingenious idea of extracting rotational energy of a rotating black hole by invoking the existence of negative energy orbits inside the ergosphere \citep{Penrose:1969:NCRS:}. He noticed that energy of a particle relative to infinity could be negative in the ergosphere where timelike Killing vector turns spacelike. Then he envisaged an infalling particle splitting into two fragments, one of which attained negative energy and fell into the hole while the other came out with energy greater than that of the incident particle. This is how the energy could be extracted from a rotating black hole. The maximum efficiency of Penrose process (PP) was limited to only $20.7 \%$ for extremely rotating black hole. In \citep{Bar-Pre-Teu:1972:APJ:,Wald:1974:PHYSR4:,Wald:1974:APJ:},  the authors independently pointed out that for the process to work relative velocity between the two fragments has to be greater than half of velocity of light. Though it was a very novel and interesting process but it was not viable astrophysically for powering the central engine of high energy objects like quasars, because there was no way to instantaneously accelerate particles to such a high relativistic velocity.

Following this, there were some variants of the process as collisional PP  were considered in \citep{Pir-Sha-Kat:1975:APJL:,Pir-Sha:1977:APJ:} and also recently in \citep{Zaslavskii:2016:PRD:,Ber-Bri-Car:2015:PRL:}. Then in 1985 Wagh, Dhurandhar and Dadhich considered magnetic version of Penrose process (MPP) \citep{Wag-Dhu-Dad:1985:APJ:} and showed that the requirement of relativistic split could be easily overcome if black hole is immersed in an external magnetic field. Now if splitting fragments have opposite charge, the required energy for one to attain negative energy could come from electromagnetic interaction thereby releasing the constraint on relative velocity altogether. It was for the first time magnetic version of the process was being considered.  The authors thus brought about astrophysical revival of the process. Further it was shown that the efficiency of the process could even exceed $100\%$ in \citep{Par-etal:1986:APJ:,Bha-Dhu-Dad:1985:JAPA:}. This was all however shown for idealized discrete particle accretion. It is however remarkable that this feature is borne out by recent fullfledged  magnetohydrodyanmical simulations \citep{Tch-Nar-McK:2011a:MNRAS:,Nar-McC:2012:MNRAS:,Narayan:2012:lecture:,Las-etal:2014:PYSR4:}. MPP has thus emerged as one of the leading mechanisms for powering the central engine of high energy sources \citep[see also][]{Williams:2004:ApJ:}.

\subsection{The formalism}

It is natural to assume that external magnetic field would also share symmetries of stationarity and axial symmetry. Assuming the field to be asymptotically uniform from the condition of electrical neutrality of the source, we thus write non-vanishing components of $4$-potential, $A^\mu$, in the form
\beq A_t = a B \left( \frac{M r}{\Sigma}
\left(1+\cos^2\theta\right) - 1 \right), \label{VecPotT}\eeq
\beq A_{\phi} = \frac{B}{2} \left(r^2+a^2-\frac{2 M r a^2}{\Sigma}
\left(1+\cos^2\theta\right)\right) \sin^2\theta,
\label{VecPotP}\eeq
where $\Sigma = r^2+a^2 \cos^2\theta$. Detailed discussion of the motion of charged particle in combined gravitational and electromagnetic fields can be found in e.g. \citep{Tur-Stu-Kol:2016:PHYSR4:}.
Note that $A_t$ vanishes on the horizon for extremal black hole which is in accordance with the well-known result of black hole physics \citep{Wald:1974:PHYSR4:,Bic-Led:2000:NCBS:} that extremal black hole, like a super conductor, expels out external fields. It is the twisting of magnetic field lines due to the frame dragging effect which
generates a quadrupole electric field indicated by $A_t$. It is this which provides energy to particle for getting onto negative energy state removing all constraints on relative velocity between fragments. Therefore it is critically  responsible for effectiveness of MPP and its eminently large efficiency. This however vanishes for extremal black hole, and hence the most favourable situation is of black hole being near extremal but not actually extremal.

The motion of charged particle is bounded by the effective potential which can be found for the equatorial radial motion as
\beq \label{Veff-mpp}
V = - q A_t - \frac{g_{t\phi}}{g_{\phi\phi}} L + \left[ \left(-\psi\right) \left(\frac{L^2}{g_{\phi\phi}}+1\right) \right]^{1/2},
\eeq
where  $L = l - q A_\phi$, $\psi = g_{tt} + \omega g_{t\phi} < 0$ and $l, q$ are particle angular momentum and charge. It is clear that $V$ can attain negative values when the term outside the radical sign is negative and dominates over the one under the radical sign for suitable set of the particle parameters. This will occur in the ergosphere bounded by static surface, $r_{\rm stat}$ from above while by horizon, $r_{+}$ from below. {Existence of negative energy state in the ergosphere is critical for MPP, and hence it has to be considered
in the equatorial plane where the largest negative energy state region is available.}

Let us now consider a neutral particle A falling on to black hole from infinity with energy $E_A \geq1$ splits into two charged fragments, B and C in the ergosphere. Without loss of generality one can put the mass of the particle A to be equal to unity. This implies, that at the point of split one can write the conservation laws in the following form
\bea E_A = m_B E_B + m_C E_C, \quad L_A = m_B L_B + m_C L_C, \\
     m_B q_B + m_C q_C = 0, \quad \dot{r}_A = m_B \dot{r}_B + m_C \dot{r}_C, \eea
where a dot denotes  derivative with respect to proper time of the corresponding particle. In order to achieve the maximum efficiency, we set $\dot{r}_B = 0$, which implies $\dot{r}_A =  m_C \dot{r}_C$.
In this case, no kinetic energy will be lost through the particle B. In addition we restrict the sum of masses of the  fragments after split to $m_B + m_C \leq 1$. Let particle B attain negative energy, $E_B<0$ and fall into the hole and particle C escapes to infinity. Then by the conservation of energy particle C comes out with energy exceeding the energy of incident particle A. This is how rotational energy of the black hole can be mined out.

%%%%%%%%%%%%%%%%%%%%%%%%%%%%%%% OBR %%%%%%%%%%%%%%%%%%%%%%%%%%%%%%%
\begin{figure}
\centering
\includegraphics[width=0.8\hsize]{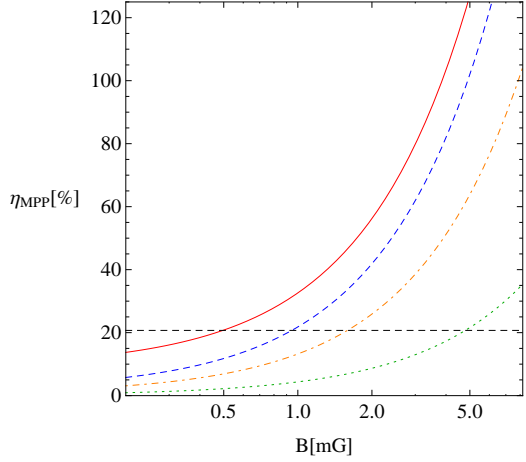}
\caption{\label{eta_rot_MPP_2} MPP efficiency is plotted against magnetic field for $a=0.1 M$ (dotted), $a=0.3 M$ (dot dashed), $a=0.5 M$ (dashed), $a=0.9 M$ (solid),  $a=M$ (horizontal dashed line), and $M = 10 M_{\odot}$. }
\end{figure}
%%%%%%%%%%%%%%%%%%%%%%%%%%%%%%% OBR %%%%%%%%%%%%%%%%%%%%%%%%%%%%%%%

\subsection{Efficiency} \label{efficiencyMPP}

In the absence of magnetic field, the efficiency of Penrose process for maximally rotating black hole   is $\eta_{\rm Kerr}^{\rm max} = (\sqrt{2} - 1)/2 \approx 20.7 \%$. We shall now show that the efficiency of MPP can increase with magnetic field arbitrarily, exceeding even $100 \%$. The efficiency is defined in the usual way as the ratio between gain and input energy.
Since particle A is neutral with $q_A=0$,  $q_B + q_C = 0$, and following \cite{Par-etal:1986:APJ:}, after several algebraic manipulations, we get
\beq
\eta_{\rm MPP} = \frac{E_C - E_A}{E_A} = \frac{-E_B}{E_A} = \chi - 1 - \frac{q_C A_t}{E_A}, \label{efficMPP}
\eeq
where
\beq \label{efficMPPimplicit}
\chi = \frac{\Omega_A - \Omega_B}{\Omega_C - \Omega_B} \frac{X_C}{X_A}, \quad X_i = g_{tt} + \Omega_i g_{t\phi}, \quad \Omega = \frac{d\phi}{dt}.
\eeq
Here subscript $i=A, B, C$ indicates the particle.
{Note that since $E_B<0$, $\eta_{\rm MPP} > 0$ indicating extraction of energy from black hole.}
 We take particle A to be neutral and assume that $E_A = 1 = m_A$. When the split occurs in close vicinity of the horizon $r = r_H$, and taking into account the effect of magnetic field (quadruple electric field) (\ref{VecPotT}) one can rewrite Eq.(\ref{efficMPP}) in explicit form as follows
\beq \label{MPPeffic}
\eta_{\rm MPP} = \frac{1}{2} \left(\sqrt{\frac{2M}{r_H}} - 1\right) + a {\cal B} \left(1 - \frac{M}{r_H}\right), \qquad {\cal B} = \frac{q_C B M}{m_C}.
\eeq
where $B$ is the strength of uniform magnetic field. The first term of Eq.(\ref{MPPeffic}) is purely geometrical and corresponds to the original Penrose process (for which maximum efficiency $\approx 20.7\%$), while the second term gives the contribution due to magnetic field. Since the dimensionless parameter ${\cal B}$ can take { any astrophysically tenable} value, the efficiency of MPP can in principle exceed $100 \%$.

In Fig.\ref{eta_rot_MPP_2} we have plotted efficiency of energy extraction for various values of magnetic field versus rotation of black hole. The efficiency keeps on increasing with rotation until extremality $a=M$ is reached, then it drops down to the pure Kerr value simply because magnetic field vanishes due to Eq.(\ref{VecPotT}) on the horizon.
%This is dictated by the well-known result \cite{Wald:1974:PHYSR4:,Bic-Led:2000:NCBS:} that no external field could be anchored on to an extremal black hole.
This is a gravitational analogue of the well-known Meissner effect in super conductor; i.e. as a conductor turns super all fields are expelled out.
Hence for MPP to work efficiently, black hole should be near  extremal but not extremal. The natural astrophysically favoured value for rotation is $a = 0.998 M$, \citep{Thorne:1974:ApJ:}.

We shall now show that how efficiency could exceed $100\%$ for astrophysically reasonable and acceptable magnetic field strength. The condition $\eta_{\rm MPP}> 100\%$ in Eq. (\ref{MPPeffic}) requires the dimensionless parameter to be ${\cal B} > 3.3$ for a particle in the vicinity of a black hole with the spin $a=0.9 M$. Inserting constants, ${\cal B} = \frac{ |q| G B M}{m c^4}$, for electron for $M=10 M_{\odot}$ we get $B \gtrsim 4$ mG for which the efficiency clearly exceeds $100\%$ (see, Fig.\ref{eta_rot_MPP_2}). {For supermassive black hole with $M=10^{9} M_{\odot}$ the corresponding magnetic field for given efficiency is $10^8$ times less than for stellar mass black hole with $M=10 M_{\odot}$.} Note that this was for the first time demonstrated for discrete particle accretion onto Kerr and Reissner-Nordstr\"om black holes respectively in \citep{Par-etal:1986:APJ:} and \citep{Bha-Dhu-Dad:1985:JAPA:}. It is remarkable that this result has withstood the detailed analysis of the process involving relativistic magnetohydrodynamical fluid accretion simulations in \citep{Nar-McC:2012:MNRAS:}. As a matter of fact the efficiency exceeding $100\%$ for mG order magnetic field marks the distinguishing signature of this process.

%%%%%%%%%%%%%%%%%%%%%%%%%%%%%%%%%%%%%%%%%%%%%%%%%%%%%%%
\section{Blandford - Znajek Mechanism} \label{sectionBZ}

In 1977 Blandford and Znajek \citep{Bla-Zna:1977:MNRAS:} proposed an interesting mechanism for extraction of rotational energy of a black hole through electromagnetic interaction. In this setting it is envisaged that magnetic field is produced by the accretion disc surrounding the hole, and magnetic field lines get twisted by the geometric frame dragging effect which then gives rise to a potential difference between the pole and equator. The discharge of which drives away energy and angular momentum \citep{Gho-Abr:1997:MNRAS:}. This is the Blandford-Znajek mechanism (BZ) and for it to work the force-free condition should be satisfied in the surrounding magnetosphere.

As in case of MPP, we assume that external field will share the stationarity and axial symmetry of the background geometry.
Skipping over details, which can be found in the literature {\citep[see, e.g. ][]{Koi-Bab:2014:APJ:}}, one can find the BZ power measured at infinity as the power of the Poynting flux
\beq \label{powerBZ}
P_{BZ} = \frac{1}{32 \Omega_H^2} \omega (\Omega_H - \omega) B_H^2 r_H^2 a^2,
\eeq
where $\Omega_H = a c/2 r_{H}$, is the angular velocity of the horizon and $\omega$ is the angular velocity of the magnetic field lines.
Note that it is the rotation of magnetic field lines that
generates the electromotive force between pole
and equator which is the driving engine for BZ.
This is how the angular velocity $\omega$ is the
key parameter and it is bounded above by the horizon
angular velocity $\Omega_H$. The extremum of Eq.(\ref{powerBZ}) gives the maximal power at $\omega = \Omega_H/2$, which corresponds to the optimal condition \citep{Bla-Zna:1977:MNRAS:}. Thus, inserting all constants explicitly we get the maximal BZ power as $P = 1.7 \times 10^{46}$ erg/s for $M= 10^9 M_\odot$ and $B=10^4$G.
The calculations of the luminosity in the framework of non-singular electrodynamics can be found in \citep{Mor-Rez-Ahm:2014:PHYSR4:}.

\subsection{Optimal regime} \label{efficiencyBZM}

{In \citep{Bla-Zna:1977:MNRAS:}, the authors introduced the quantity $\epsilon$ defined as the ratio between the output energy from the black hole and the rotational energy of a black hole. It is different from the efficiency defined in MPP section, however the following discussion will be useful for the estimates of BZ in optimal regime, i.e. in a regime with maximum output power.} The rate of the energy outflow and the rotational energy loss rate are given by
\beq
d P_{BZ} =  -\frac{1}{2 \pi} \omega I d \Psi, \qquad d P_{\rm rot} =  -\frac{1}{2 \pi} \Omega_H I d \Psi, - q A_t - \frac{g_{t\phi}}{g_{\phi\phi}} L
\label{PBZ-bz-rot}
\eeq
where  $L$ is as defined in Eq (4), and  $\Psi$ and $I$ are the constants along the magnetic surface, corresponding to the magnetic flux and outward current, respectively. Without loss of generality we use the units where $M=1$. We will however restore $M$ at the end of this section for estimates. Thus, the efficiency as defined by \citep{Bla-Zna:1977:MNRAS:} takes the form
\beq
\epsilon_{\rm BZ} = \frac{P_{\rm BZ}}{P_{\rm rot}} = \frac{\omega}{\Omega_H}.
\eeq
The maximum of 100\% is unattainable, because in that case the BZ power tends to zero according to (\ref{powerBZ}).
This implies that the condition for the BZ to work is governed by inequality
\beq \label{omlessom}
\omega < \Omega_H.
\eeq
It is easy to see from (\ref{powerBZ}) that the optimal case corresponding to the maximal power with BZ energy extraction is given by $\omega = \Omega_H / 2$.
In that case $\epsilon_{\rm BZ} = 0.5$, while larger $\epsilon_{\rm BZ}$ would decrease the power of BZ.
Note, that the expression for $\epsilon_{\rm BZ}$, which was introduced in \citep{Bla-Zna:1977:MNRAS:}, does not depend explicitly on the strength of magnetic field. So, we find it in a different way as an explicit ratio of the actual extracted energy and the maximum extractable energy.
The energy transferred from black hole horizon and measured at infinity, can be found in the form \citep{Koi-Bab:2014:APJ:}
\beq \label{Eneg-bz}
E_{\rm BZ} = \frac{1}{2} \frac{\alpha^2 + 2 g_{\phi \phi} \omega (\omega-\Omega_H)}{\alpha^2 + g_{\phi \phi} (\omega-\Omega_H)^2} \alpha {B}^2,
\eeq
where $\alpha^2 = -1/g^{tt}$. Close to the horizon, $\alpha \rightarrow 0$, which implies that the transfer of negative energy is realized when condition (\ref{omlessom}) holds. {Thus, the condition $\omega < \Omega_H$ also corresponds to the inflow of negative energy towards the black hole. } Though matter can never escape the horizon of a black hole, the statement above can be interpreted as transport of negative electromagnetic energy from infinity to the horizon  \citep[see, details in][]{Koi-Bab:2014:APJ:}. The deposition of negative energy on black hole implies that energy is extracted from the hole.
The condition (\ref{omlessom}) for BZ plays similar role as the condition $V<0$ from (\ref{Veff-mpp}) plays for MPP.  Dividing energy (\ref{Eneg-bz}) by $E_{\rm rot}$, given by Eq.(\ref{ExtRotEn}), we get $\epsilon_{\rm BZ}$ in a form proportional to the square of magnetic field:
\beq
\epsilon_{\rm BZ} = \frac{\sqrt{2}  F ~B^2}{F+24 \Delta  r^2} \left(\frac{r\Delta}{a^2 (r+2)+r^3}\right)^{1/2} \frac{1}{\sqrt{2}-\sqrt{A+1}},
\eeq
where  $F=a^6 (r+2)^2+2 a^4 r^3 (r+2)+a^2 r^4 \left(r^2-8\right)-8 (r-2) r^5$, and $A = \sqrt{1 - a^2}$.

Let us estimate the strength of magnetic field required to power BZ at optimal regime, i.e. when $\epsilon_{\rm BZ} = 0.5$. The energy available for the extraction from the black hole with the spin $a=0.9M$ is $E_{\rm rot} = 0.15 M c^2$ according to Eq.(\ref{ExtRotEn}). Computing the output energy at the point where (\ref{Eneg-bz}) is maximal with $a=0.9M$, we get BZ energy as $E_{\rm BZ} = 0.11 B^2 r_H^3$. Thus, the magnetic field required to power BZ at optimal regime can be estimated as
\beq \label{BB22}
B \approx 6.3 \times 10^8 \left(\frac{a}{M} \right) \left(\frac{10^9 M_{\odot}}{M}\right) {\rm G}.
\eeq
Thus, in contrast to the MPP which is highly efficient even for mG magnetic field, the BZ requires magnetic field $\gtrsim 10^5$G for it to be even operative. This is what we compute next.

\subsection{Threshold magnetic field} \label{seclimitB}

{ For BZ mechanism to be operative, there must exist a force-free magnetosphere around a rotating black hole.}
Let us discuss now the mechanism which could support the existence of the force-free magnetosphere. According to \citep{Wald:1984:book:}, for the Kerr black hole immersed in asymptotically uniform magnetic field $B_0$ the total flux through the upper half of the horizon is $\Phi_H = 4\pi B_0 M (r_H - M)$.
This implies that the voltage difference between the horizon and infinity $\Delta V \sim \Omega_{H} \Phi / 2\pi$ for the background parameters used in the previous subsection is of the order
\beq \Delta V \sim 10^{12} \left(\frac{a}{M}\right) \left(\frac{M}{10 M_\odot}\right) \left(\frac{B_0}{10^{4} \rm{G}}\right) {\rm V}, \eeq
or $10^8$ times larger for supermassive black holes with $M=10^9 M_{\odot}$.
Such a huge voltage drop along the field lines accelerates the stray electrons moving in the black hole vicinity to large values of Lorentz $\gamma$ factor and, upon colliding with stray photons or charges of opposite sign, they would produce a cascade
of electron-positron pairs. Very quickly therefore, the vacuum surrounding the black hole would be filled with a highly conducting plasma.
%The plasma-loaded field lines can serve as conducting wires to complete the electric circuit between the pole and equator of the horizon.
This would require a critical threshold magnetic field strength for creation of cascade of electron-positron pairs for producing environment of force free conducting plasma.
The approximate estimation of the threshold magnetic field to create the electron-positron pairs near black hole horizon is given by \cite{Bla-Zna:1977:MNRAS:}
\beq
B_c \approx 6.2 \times 10^4 \left(\frac{M}{a}\right)^{\frac34} \left(\frac{10 M_\odot}{M}\right)^{\frac12} {\rm G}. \label{critB}
\eeq
This yields the threshold limit of the order $10^4$G where we have assumed that other contributions due to gyration, Doppler effect and Compton scattering are negligible. It also corresponds to the condition of nearly infinite conductivity along the magnetic field lines, which is required to power BZ. {For supermassive black holes the threshold magnetic field is of the order of tens of Gauss.} It is remarkable that MPP could have more than $100\%$ efficiency for a milli Gauss magnetic field for which BZ is entirely non-operative. %This is a clear distinction between the two processes.

\section{Conclusion} \label{SecSummary}

It was in 1969 that Penrose proposed an ingenious process of energy extraction \citep{Penrose:1969:NCRS:} from a rotating black hole by invoking a remarkable property of Kerr geometry that a particle can have total energy negative relative to an observer at infinity for suitable particle parameters. This was a very novel and
remarkable process which was entirely powered by the Kerr geometry, and it could very well fit in as an energy source for then recently discovered bizarre high energy objects, quasars. However it was soon realized that the process could not be efficient enough to foot the bill \citep{Bar-Pre-Teu:1972:APJ:,Wald:1974:APJ:}. Then in 1977 came in BZ \citep{Bla-Zna:1977:MNRAS:} in which a rotating black hole was sitting in a magnetic field and the negative energy states available near black hole horizon facilitated creation of potential difference between the pole and equator. The discharge of which drove energy and angular momentum away thereby extracting rotational energy of black hole.

In 1985 magnetic version of PP was considered \citep{Wag-Dhu-Dad:1985:APJ:} and it turned out to be enormously efficient, so much so that its efficiency could exceed $100\%$ \citep{Par-etal:1986:APJ:,Bha-Dhu-Dad:1985:JAPA:}  {even for as low a magnetic field as a milli Gauss}. This was a clearcut signature of MPP that distinguishes it from all other mechanisms. This was a prediction based on discrete particle accretion which had wonderfully stood the test of the recent numerical simulation models employing fully relativistic magnetohydrodynamic fluid accretion \citep{Nar-McC-Tch:book:2014:,Las-etal:2014:PYSR4:}. It has been shown that MPP is indeed the most successful and promising mechanism for powering high energy sources, quasars and AGNs. 

  As argued above BZ is really a high magnetic field limit of MPP;i.e. the two are one and the same process -- BZ operates only on high magnetic field regime. The extracted energy is essentially given by $ - q A_t + \omega (l - q A_\phi) < 0 $ where $\omega = -g_{t\phi}/g_{\phi\phi}$. Note that electric potential $A_t$ is produced by twisting of magnetic field lines which is caused by black hole rotation, and hence the source of extracted energy is in essence rotation and not electromagnetic. Magentic field is a catalytic mediator that facilitates the process of extraction of rotational energy of black hole electromagnetically.

BZ requires high threshold magnetic field for ionization of vacuum so as to form force free magnetosphere. On the other hand for MPP it is envisaged that such an ionized environment generally obtains in  magnetized accretion disks \citep{Penna-etal:2010:MNRAS:} around rotating black holes. Given that, the process works super efficiently for as low as milli Gauss field for stellar mass black hole. Magnetic fields around stellar or supermassive black holes could range from very low in milli Gauss to very high $10^{6-8}G$, and for AGNs it could go down to micro Gauss. MPP would however cover this entire range tending to BZ at high end. The point we would like to make is that energy extraction by MPP could be very efficient even for stellar mass black hole with low magnetic field. The MHD version of MPP with low magnetic field should be studied and we believe that it would still bear out the promise of efficiency $>100\%$. This is a clear and definitive prediction which should be tested. That is,  in low magnetic field environment of AGNs and supermassive black holes should be simulated for relativistic magnetohydrodynamic flow and see whether the prediction is borne out or not?

% ====================================== %
% ====================================== %
% ====================================== %
% ====================================== %

\section*{Acknowledgments}

The authors would like to acknowledge the
Institutional support of the Faculty of Philosophy and Science of
the Silesian University in Opava.
N.D. thanks Albert Einstein Institute, Golm for a summer
visit.
A.T. acknowledges the Czech Science Foundation Grant No. 16-03564Y and the internal student grant of
the Silesian University SGS/14/2016.
B. A. acknowledges
Inter-University Centre for Astronomy and Astrophysics,
Pune, India, and Goethe University, Frankfurt am Main,
Germany, for warm hospitality.
This research is supported in part by Projects No. VA-FA-F-2-008 and
No.YFA-Ftech-2018-8 of the Uzbekistan Ministry
for Innovation Development,
and by the Abdus Salam International Centre for
Theoretical Physics through Grant No. OEA-NT-01 and
by an Erasmus+ exchange grant
between SU and NUUz.
Z.S. acknowledges the
Albert Einstein Centre for Gravitation and Astrophysics
supported by the Czech Science Foundation Grant No.
14-37086G.

%%%%%%%%%%%%%%%%%%%%%%%%%%%%%%%%%%%%%%%%%%%%%%%% BIB %%%%%%%%%%%%%%%%%%%%%%%%%%%%%%%%%%%%%%%%%%%%%%%%

\def\prc{Phys. Rev. C}
\def\pre{Phys. Rev. E}
\def\prd{Phys. Rev. D}
\def\jcap{Journal of Cosmology and Astroparticle Physics}
\def\apss{Astrophysics and Space Science}
\def\mnras{Monthly Notices of the Royal Astronomical Society}
\def\apj{The Astrophysical Journal}
\def\aap{Astronomy and Astrophysics}
\def\actaa{Acta Astronomica}
\def\pasj{Publications of the Astronomical Society of Japan}
\def\apjl{Astrophysical Journal Letters}
\def\pasa{Publications Astronomical Society of Australia}
\def\nat{Nature}
\def\physrep{Physics Reports}
\def\araa{Annual Review of Astronomy and Astrophysics}
\def\apjs{The Astrophysical Journal Supplement}
\def\aapr{The Astronomy and Astrophysics Review}

\bibliographystyle{mnras}

%\input{refdef}
%\bibliography{reference}

\end{document}